\documentclass[twocolumn,showpacs,preprintnumbers,amsmath,amssymb]{revtex4}

\usepackage{graphicx}
\usepackage{dcolumn}
\usepackage{bm}

\begin{document}

\input{epsf}
\preprint{}

\title{Thermodynamics of the half-filled Kondo lattice model around the atomic 
limit}

\author{Qiang Gu}

\affiliation{Max-Planck-Institut f\"{u}r Physik komplexer Systeme, 
N\"{o}thnitzer Stra{\ss}e 38, 01187 Dresden, Germany}

\date{\today}

\begin{abstract}
We present a perturbation theory for studying thermodynamic properties of the 
Kondo spin liquid phase of the half-filled Kondo lattice model. The grand 
partition function is derived to calculate chemical potential, spin and 
charge susceptibilities and specific heat. The treatment is applicable to the 
model with strong couplings in any dimensions (one, two and three dimensions). 
The chemical potential equals zero at any temperatures, satisfying the 
requirement of the particle-hole symmetry. Thermally activated behaviors of 
the spin(charge) susceptibility due to the spin(quasiparticle) gap can be seen 
and the two-peak structure of the specific heat is obtained. The same treatment 
to the periodic Anderson model around atomic limit is also briefly discussed.
\end{abstract}


\maketitle

The Kondo lattice model (KLM) and the periodic Anderson model (PAM) are two 
prototype models for the heavy fermion system and related 
materials\cite{lee,fulde,hewso}. Both models describe a band of itinerant 
conduction electrons interacting with a lattice of magnetic impurities. At 
half-filling, they are usually viewed as standard models to understand the 
so-called Kondo insulator. The PAM is written as
\begin{eqnarray}
H_{PAM} &=& -t \sum_{\langle i,j\rangle,s} (c^{\dag}_{is}c_{js}+H.c.)
  + V\sum_{i,s}( c^{\dag}_{is}f_{is}+H.c.) \nonumber \\
  &&+ \epsilon_f \sum_{i}n^f_i+U_f\sum_{i} n^f_{i\uparrow}n^f_{i\downarrow} ,
\end{eqnarray}
where the fermion operators $c^{\dag}_{is}$ ($f^{\dag}_{is}$) create conduction
(impurity) electrons and the summation $\sum_{\langle i,j\rangle}$ is taken over 
all nearest neighbors. In the Kondo limit, $U_f \gg V$ and $\epsilon_F \gg 
\epsilon_f$ ($\epsilon_F$ is the Fermi energy), the charge fluctuation of 
$f$ electrons is suppressed and each impurity site is occupied by one and only 
one $f$ electron. Then the PAM can be simplified and reduced to the 
KLM\cite{schr},
\begin{equation}
H_{KLM} = -t \sum_{\langle i,j\rangle,s} (c^{\dag}_{is}c_{js}+H.c.)+
 J \sum_{i}{\bf S}_{i}\cdot {\bf s}^c_{i} ,
\end{equation}
with the spin-exchange coupling $J=8V^2/U_f$ under symmetric conditions. 
Here ${\bf s}^c=\sum_{s,s'} \frac 12 {\bf \sigma}_{s,s'}c^\dag_{is}c_{is'}$ 
is the spin-density operator of conduction electrons and 
${\bf S}=\sum_{s,s'} \frac 12 {\bf \sigma}_{s,s'}f^\dag_{is}f_{is'}$ denotes 
the localized spin with $f^\dag_{is}$ ($f_{is}$) being the pseudo-fermion 
operators. ${\bf \sigma}_{s,s'}$ is the Pauli matrix. 

The purpose of this work is mainly to consider the half-filled KLM whose 
ground state and finite-temperature properties have been intensively studied in 
recent years\cite{tsun,wang,shi,capp,shib}. The ground-state 
phase diagram has been now well established\cite{tsun}. Owing to the Kondo 
screening effect, the one dimensional (1D) model has a Kondo spin liquid (KSL) 
state at all values of the coupling $J/t$. In higher dimensions, it is 
suggested that competition between the Kondo screening and the 
Ruderman-Kittel-Kasuya-Yosida interaction leads to quantum phase transitions 
from the KSL state to the antiferromagnetically ordered state as $J/t$ 
decreases, with the critical point $(J/t)_c \approx 1.4$ for two-dimensional 
(2D) and $2.0$ for three-dimensional (3D) KLM's\cite{wang,shi,capp}. The KSL 
phase is characterized by three different energy scales: the spin ($\Delta_s$), 
charge ($\Delta_c$) and quasiparticle ($\Delta_{qp}$) gaps. The spin gap 
vanishes at the critical point while the other two gaps still remain finite in 
the ordered phase. The finite-temperature properties of the KSL phase are 
dominated by all three energy scales. A number of powerful numerical methods 
including quantum Monte Carlo (QMC) simulations\cite{capp}, the density matrix 
renormalization group method\cite{shib} and the finite-temperature Lanczos 
technique\cite{haule} have been used to investigate the thermodynamic 
properties of the KLM in 1D and 2D. However, less analytical work is known, 
especially for the higher dimensional models.

In this paper, we propose a finite-temperature perturbation theory to study the 
thermodynamics of the KSL phase around the atomic limit $t\to 0$. 
We believe that the approach is applicable to the KLM in any dimensions and at 
any temperatures as long as the coupling is sufficiently strong, $J\gg t$. 
The model under consideration is a generalized KLM by introducing an on-site 
Coulomb interaction between conduction electrons to the Hamiltonian (2),
\begin{equation}
H = H_{KLM} + U_c\sum_{i} n^c_{i\uparrow}n^c_{i\downarrow} .
\end{equation}
The hopping term in Eq. (2) is treated as a perturbation. A similar treatment has 
been applied to pure spin systems with the dimerized ground state\cite{gu} and 
yields good agreement with experiment results and QMC simulations\cite{john}. 

\begin{table}
\caption{\label{tab:table1}Eigen states and energies of each atom. $h$ is the 
external magnetic field. $\alpha=1,0,-1$ is the spin-$z$ component of triplet 
excitations and $s=\frac 12$ ($-\frac 12$) corresponds to spin up (spin 
down) of hole and double occupancy states. }
\begin{ruledtabular}
\begin{tabular}{cccc}
 Eigenstate & Spin index & Eigenvalue & Electron numbers\\
\hline
$|s_i\rangle$           & 0        & $-\frac 34 J$ &1 \\
$|t_i^{\alpha}\rangle$& $\alpha$ & $\frac 14 J -\alpha h$  &1 \\
$|a_{is}\rangle$  & $s$ & $-s h$   &0 \\
$|b_{is}\rangle$  & $s$ & $U_c-s h$  &2 \\
\end{tabular}
\end{ruledtabular}
\end{table}

The strong coupling limit has been taken as a good starting point to understand 
the ground state properties of the KSL phase. Various techniques based 
on this limit, such as direct strong-coupling expansions\cite{tsun,shi}, slave 
particle approach\cite{sigr,jure} and projector techniques\cite{eder}, are 
employed to calculate the ground state energy and the excitation spectra, and 
even to determine the quantum transition point of higher-dimensional KLM's. In 
the limit case, the ground state and low-lying excited states can be described 
by simple wave functions and thus it is much easier to define the energy 
gaps\cite{tsun}. For each atom, the Hilbert space consists of 
eight possible quantum states (see table I): a singlet state 
$|s_i\rangle=\frac 1{\sqrt{2}}(c^{\dag}_{i\uparrow}f^{\dag}_{i\downarrow} - 
c^{\dag}_{i\downarrow}f^{\dag}_{i\uparrow})|0\rangle$, three degenerate triplet 
states $|t_i^1\rangle=c^{\dag}_{i\uparrow}f^{\dag}_{i\uparrow}|0\rangle$, 
$|t_i^{-1}\rangle=c^{\dag}_{i\downarrow}f^{\dag}_{i\downarrow}|0\rangle$ and 
$|t_i^{0}\rangle=\frac 1{\sqrt{2}}( c^{\dag}_{i\uparrow}f^{\dag}_{i\downarrow} + 
c^{\dag}_{i\downarrow}f^{\dag}_{i\uparrow})|0\rangle$, two hole states 
$|a_{is}\rangle=f^{\dag}_{is}|0\rangle$ and two double occupancy states 
$|b_{is}\rangle=c^{\dag}_{i\uparrow}c^{\dag}_{i\downarrow}f^{\dag}_{is}|0\rangle$. 
The unperturbed lattice eigen states are simply the direct products of the atomic 
states, e.g., $|s_1s_2...s_N\rangle$ is the ground state. The lowest spin 
excitation is promoting one singlet to triplet. And the lowest charge excitation 
is to create a pair of one vacant and one doubly occupied site, with a 
spin-singlet combination of the two localized spins. The quasiparticle is defined 
as either one vacancy or one double-occupancy configuration. One can get a 
qualitatively correct picture of the KSL phase for smaller couplings from the 
limit case. Naturally, we expect that it is also an appropriate starting point for 
understanding thermodynamic properties.

We use the grand canonical ensemble. First the grand partition function is 
calculated, $Z(\beta,\mu) = Tr [exp(-\beta (H-\mu n^c))]$ 
where $\beta=1/{k_BT}$, $\mu$ is the chemical potential and $n^c$ is the 
number operator of conduction electrons. Since the $f$ electrons are completely 
localized and they have no contributions to the fluctuation of the electron number 
in the KLM, only the number of conduction electrons is considered. Performing the 
cumulant expansion, the partition function reads
\begin{equation}
Z(\beta,\mu) = Z_0 exp[ \sum_{n=1}^{\infty} U_n ],
\end{equation}
with the cluster function $U_n$ defined as 
\begin{eqnarray}
U_1 &=& \frac {Z_1}{Z_0},  \nonumber \\
U_2 &=& \frac {Z_2}{Z_0} - \frac12 \left( \frac {Z_1}{Z_0} \right)^2, 
\nonumber \\
U_3 &=& \frac {Z_3}{Z_0} - \frac {Z_2 Z_1}{Z_0^2} + \frac 13 \left( \frac {%
Z_1}{Z_0} \right)^3,...
\end{eqnarray}
Here $Z_n$ is the $n$-th order perturbation expansion of the partition 
function, 
\begin{eqnarray}
&&Z_n = {(-1)}^n \int\limits_{0}^{\beta}d\tau_1 
 \int\limits_0^{\tau_1}d\tau_2...\nonumber\\
 &&\times \int\limits_0^{\tau_{n-1}}d\tau_n Tr [
e^{-\beta (H^0-\mu n^c)} H^I(\tau_1)H^I(\tau_2)...H^I(\tau_n) ] ~
\end{eqnarray}
and $H^I(\tau) = e^{\tau (H^0-\mu n^c)} H^I e^{-\tau (H^0-\mu n^c)}$. Here 
$Z_n$ can be integrated out in the basic set of eigenstates of the unperturbed 
Hamiltonian\cite{gu}. It contains the product of $n$ perturbation matrices, 
\begin{widetext}
$$(H^I)_{nk}(H^I)_{kl}...(H^I)_{mn}=\sum_{\langle i_1j_1\rangle}
\sum_{\langle i_2j_2\rangle}...\sum_{\langle i_nj_n\rangle}
(H^I_{i_1j_1})_{nk}(H^I_{i_2j_2})_{kl}...(H^I_{i_nj_n})_{mn}.$$
\end{widetext}
$(H^I_{ij})_{mn}=-t\langle m|\sum_s (c^{\dag}_{i,s}c_{j,s}+H.c.)|n\rangle$ 
is the matrix element of the perturbation in the Hilbert space of the atom 
lattice. Nevertheless only those terms in which all the sites 
$i_1,j_1;i_2,j_2;...;i_n,j_n$ are connected by the interaction are taken into 
account in the cluster function $U_n$. 

The computation of the perturbation expansions is straightforward but very 
laborious. Above all, one has to compute the perturbation matrix elements 
$(H^I_{ij})_{mn}$. It is important to note that both $f_s$ and $c_s$ are 
fermionic operators and the sign changes when they exchange their sites. For 
example,
\begin{eqnarray}
H^I_{ij}|a_{i\uparrow}t_{j}^{+1}\rangle &=& t|t_{i}^{+1}a_{j\uparrow}\rangle, 
\nonumber \\
H^I_{ij}|b_{i\uparrow}t_{j}^{+1}\rangle &=& -t|t_{i}^{+1}b_{j\uparrow}\rangle.
\end{eqnarray}
$|a_{is}\rangle$ and $|b_{is}\rangle$ are fermionlike states, while 
$|t_i^\alpha \rangle$ and $|s_i\rangle$ are bosonlike states, 
\begin{eqnarray}
|a_{is}b_{js'}\rangle &=& -|b_{js'}a_{is}\rangle,\nonumber \\
|s_it_j^\alpha\rangle &=& |t_j^\alpha s_i\rangle.
\end{eqnarray}
${(H^I_{ij})_{mn}}$ is a real, diagonal matrix. For each nonzero element, the 
final and initial states $\langle m|$ and $|n\rangle$ differ only in the states 
of the two adjacent atoms and the rest of the lattice is unchanged. So only 
$8\times 8$ different states are concerned. The interaction term just transfers 
an electron from one site to its neighbors and so it makes the states of the two 
involved sites definitely changed but conserves the number of electrons and the 
$z$ component of spins. Taking advantage of this point, the calculation of the 
matrix elements and the cluster functions can be greatly simplified. For example, 
one can easily get that all diagonal elements of perturbation matrix are zero. 
To the {\sl d}-dimensional cubic lattice KLM, all the odd-order cluster functions 
also vanish. In this paper, we only evaluate the second-order expansion ($n=2$).

Before calculating thermodynamic properties, one must first determine the 
chemical potential $\mu$ by solving the equation 
\begin{equation}
\langle n^c\rangle =-\frac 1N \frac {\partial F(\beta,\mu)}{\partial \mu},
\end{equation}
where $\langle n^c\rangle$ is the average number of conduction elections per site, 
$N$ is the number of sites and $F(\beta,\mu)= - \frac 1{\beta} ln[Z(\beta,\mu)]$ is 
the free energy. Generally, the chemical potential is a function of the temperature. 
Nevertheless, $\mu$ should be exactly equal to zero at any temperatures due to the 
particle-hole symmetry at half filling. There is one conduction electron 
per site on average in this case. Setting $\langle n^c\rangle=1$ in the above equation, 
we obtain that the requirement is well satisfied up to the second-order expansion.  
 
\begin{figure}
\center{\epsfxsize=75mm \epsfysize=85mm \epsfbox{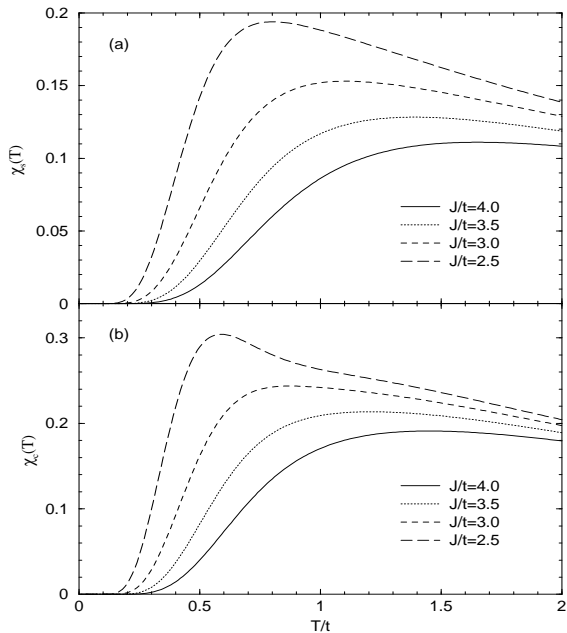}}
\caption{\label{fig:epsart} Spin (a) and charge (b) susceptibilities $\chi_s$, 
$\chi_c$ vs $T/t$ for the 1D KLM with $U_c=0$.}
\end{figure}

We proceed to calculate the spin susceptibility 
$\chi_s (T) = -\frac 1N {(\frac{\partial^2 F}{\partial h^2})_{h=0}}$, charge 
susceptibility $\chi_c (T) = - \frac{\partial n^c}{\partial \mu}$ and specific heat 
$C(T) = -\frac 1N \beta^2 \frac{\partial ^2(\beta F)}{(\partial \beta)^2}$. All the 
thermodynamic quantities are derived from the free energy by a standard procedure. 
Fig. (1) shows numerical results of spin and charge susceptibilities for the 
1D KLM. Raising temperature from zero, both $\chi_s(T)$ and $\chi_c(T)$ rise 
exponentially, indicating the existence of the spin and quasiparticle gaps, 
and then they reach a maximum. The peaks of $\chi_s (T)$ and $\chi_c (T)$ move 
to lower temperatures with $J/t$ decreasing, suggesting that the $\Delta_s$ 
and $\Delta_{qp}$ diminish correspondingly. As discussed by 
Shibata {\sl et al.}\cite{shib} and Haule {\sl et al.}\cite{haule}, both 
$\Delta_s$ and $\Delta_{qp}$ are responsible for the spin susceptibility but 
mainly the lower one dominates the low temperature behaviors. In contrast, 
$\chi_c$ is governed by $\Delta_{qp}$ only. Since $\Delta_{qp}<\Delta_s$ near 
the atomic limit, both $\chi_s (T)$ and $\chi_c (T)$ are dominated by the 
quasiparticle gap. We find that the activation energies estimated by fitting 
the spin and charge susceptibilities are similar when $J/t>4$, consistent with 
the conclusion of the numerical results\cite{shib,haule}. However the 
second-order perturbation result is not accurate enough to discuss the 
delicate relation between the activation energies of $\chi_s (T)$ and 
$\chi_c (T)$ and the spin and quasiparticle gaps for smaller couplings.

\begin{figure}
\center{\epsfxsize=75mm \epsfysize=85mm \epsfbox{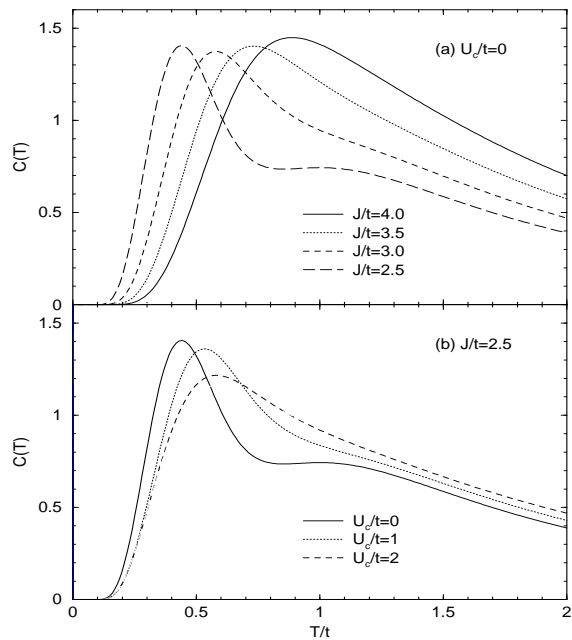}}
\caption{\label{fig:epsart} Specific heat $C(T)$ vs $T/t$ of the 1D KLM with 
(a) $U=0$ and $J/t=4.0, 3.5, 3.0, 2.5$; (b) $J/t=2.5$ and $U_c=0, 1.0, 2.5$.}
\end{figure}

For $J/t\lesssim 2.5$, there appears an unphysical protrusion on the peak of 
$\chi_c (T)$. This value is an approximate lower limit of the $J/t$ range for 
which the perturbation theory is reasonable for all temperatures. 
Within the second-order perturbation theory, the lower limit of $J/t$ is about 
$2.5$,  $3.0$ and $4.0$ for 1D, 2D and 3D KLM's respectively. This is consistent
with the zero-temperature strong-coupling expansions by which the spin gap within 
second order is given as $\Delta^{(2)}_s=J-\frac {20dt^2}{3J}$\cite{tsun}. 
$\Delta^{(2)}_s$ tends to zero quickly at $J/t \approx 2.58, 3.65$ and 
$4.47$ for $d=1, 2$ and $3$. Considering higher-order expansions is expected 
not only to improve the accuracy of the result, also to reduce the lower limit 
of $J/t$. This task might be achieved by the high order series expansion 
technique\cite{gelf} which has been successful in explaining 
ground-state properties of KLM\cite{shi}. The methodology for carrying out high 
order expansions at finite temperature for spin systems has been developed 
recently\cite{elst}. 

\begin{figure}
\center{\epsfxsize=75mm \epsfysize=85mm \epsfbox{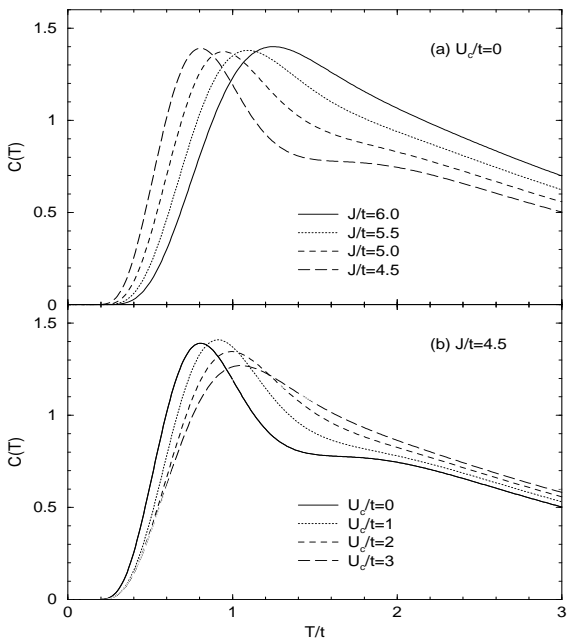}}
\caption{\label{fig:epsart} Specific heat of the 3D KLM. }
\end{figure}

Results of the specific heat for 1D and 3D KLM's are presented in Figs. (2) and 
(3). The specific heat contains information on both spin and charge degrees 
of freedom. The charge fluctuation originates from the movement of 
conduction electrons. Moving an electron from one site to its neighbors on the 
ground state background creates a pair of one vacant and one doubly occupied 
site. The energy cost just is the charge gap. The charge fluctuation is greatly 
suppressed due to the large charge gap near the atomic limit, so only one peak 
attributed to the spin excitations can be seen. The location of the peak 
depends on the spin gap which varies with $J$. The hoping term stirs up the 
charge fluctuations. Then a new peak originating from the charge degree of freedom 
begins to be visible as $J/t$ is reduced to $3.0$, as shown in Fig. 2(a). The
new peak corresponds to the specific heat of the free conduction electrons and
is independent of $J$. The on-site Coulomb repulsion in the conduction electron 
band will enhance the charge gap and so it suppresses the charge fluctuation. 
As shown in Fig. 2(b), the second peak in $C(T)$ will vanish with increasing 
$U_c$. Comparing Figs. (3) and (2), the results for the 3D KLM exhibit similar
features to those for the 1D case. As the $T=0$ perturbation theory shows, 
various properties, including the characteristic energy gaps, of the KSL phase are 
mainly determined by the local dimers in the atomic limit\cite{tsun}, so the 
dimensionality does not make much difference in thermodynamic properties. This 
point is also indicated in numerical results for the 2D KLM\cite{haule}.

At last, we notice that the above treatment can be also applied to deal with the 
PAM, but with more complexity. Contrary to the KLM, the charge degrees of freedom 
of $f$ electrons must be considered in the PAM. The vacancy and double occupancy 
of the impurity site are possible and thus the Hilbert space of each atom is 
greatly enlarged. In general case each atom has 16 quantum states. The atomic limit 
of the PAM has been investigated by many authors\cite{alas,manc,noce,long}. 
Choosing the atom Hamiltonian as the unperturbed Hamiltonian, we can evaluate the
thermodynamics near the atomic limit by perturbation theory as above. The physical 
properties of the PAM are more complicated than those of the KLM. For example, 
the specific heat will exhibit two peaks even in the atomic limit\cite{bern}. 
More recently, Moskalenko {\sl et al.} calculated the one-particle Green's function 
of the PAM by the hopping perturbation treatment\cite{mosk}. 
 
In summary, we have studied thermodynamic properties of the Kondo spin liquid 
phase of the half-filled Kondo lattice model by a finite-temperature 
perturbation theory. The chemical potential, spin and charge susceptibilities 
and specific heat are calculated. The results are consistent qualitatively 
with those obtained by numerical methods. It proves that the strong coupling 
limit is a reasonable starting point to study thermodynamics of the Kondo spin 
liquid phase. The accuracy of obtained results is expected to be improved by 
taking higher-order expansions into account. 

The author would like to thank P. Thalmeier for helpful discussions.

\appendix
\section{}

The free energy per site within second-order perturbation is given as
$$ f(\beta,h)=-\frac 1{\beta} \left[ ln(z_0) + d \frac {z_2}{z_0^2} \right] ,$$
where $d=1,2,3$ is the dimensionality and
\begin{widetext}
\begin{eqnarray}
z_0 &=& \left[ 1+ e^{-\beta(U_c-2\mu n^c)} \right]f(h) + e^{-\beta(-\frac {3J}4 -\mu n^c)} 
  + e^{-\beta(\frac J4-\mu n^c)}\left[ 1+ f(2h) \right] ,\nonumber 
\end{eqnarray}
\begin{eqnarray}
z_2 &=& \frac {\beta^2 t^2}4 \left[ e^{-\beta(-\frac {3J}4-\mu n^c)} + 
    e^{-\beta(-\frac {3J}4 + U_c - 3\mu n^c)} \right] f(h) 
    + \frac {5\beta^2 t^2}4 \left[ e^{-\beta(\frac J4-\mu n^c)} + 
    e^{-\beta(\frac J4 +U_c-3\mu n^c)}\right] f(h) \nonumber \\
  &+& \beta^2 t^2 \left[ e^{-\beta(\frac J4-\mu n^c)} + 
    e^{-\beta(\frac J4 + U_c - 3\mu n^c)}\right] f(3h) \nonumber \\
  &+& \frac {3\beta t^2}{2J}\left[ e^{-\beta(-\frac {3J}4-\mu n^c)} +  
    e^{-\beta(-\frac {3J}4+U_c-3\mu n^c)} - e^{-\beta(\frac J4-\mu n^c)} -  
    e^{-\beta(\frac J4+U_c-3\mu n^c)} \right]f(h) \nonumber \\
  &-& \frac {\beta t^2}{J-2U_c} \left[ 4 e^{-\beta(\frac J2-2\mu n^c)}f(2h) +
    10 e^{-\beta(\frac J2-2\mu n^c)} - 10 e^{-\beta(U_c-2\mu n^c)} \right] \nonumber \\
  &+& \frac {4\beta t^2}{J+2U_c} \left[ e^{-\beta(-\frac J2-2\mu n^c)}f(2h) +
    e^{-\beta(-\frac J2-2\mu n^c)} - e^{-\beta(U_c-2\mu n^c)} \right] \nonumber \\
  &+& \frac {2\beta t^2}{3J+2U_c} \left[ e^{-\beta(-\frac {3J}2-2\mu n^c)} -
    e^{-\beta(U_c-2\mu n^c)} \right] + 
    \frac{16\beta t^2 U_c}{J^2-4U_c^2} e^{-\beta(U_c-2\mu n^2)} f(2h) \nonumber 
\end{eqnarray}
\end{widetext}
with $f(h)=e^{-\frac 12\beta h}+e^{\frac 12\beta h}$.

\end{document}